\documentstyle [12pt]{article}
\advance\voffset by -1 in
\advance\hoffset by -.5 in
\def\d{\delta}

\def\p{\partial}

\def\O{\Omega}
\def\Om0{\Omega^0}
\def\O1{\Omega^1}

\def\tf{\tilde{f}}

\def\res{{\rm res}\:}

\begin{document}

\centerline{{\Large\bf On the constrained KP hierarchy II}}
\centerline{An additional remark}
\vspace{.4in}
\centerline{{\bf L.A.Dickey}}

\vspace{.2in}
\centerline{Dept. Math., University of Oklahoma, Norman, OK 73019\footnote{
Internet: "ldickey@nsfuvax.math.uoknor.edu"}}

\bigskip
\centerline{{\bf Abstract}}

\vspace{.1in}
\small
This is an additional remark to the paper (hep-th 9411005) concerning a
Hamiltonian structure of suggested there system of equations. The remark is
inspired by a letter from L. Feher and I. Marshall.
\normalsize

\vspace{.3in}
After this paper had appeared in hep-th (hep-th 9411005) I had a letter from L.
Feher and I. Marshall to the effect that they probably knew the answer to one
of the questions I put at the end of the article. Namely, the system of
equations (3) must be Hamiltonian with respect to the direct sum of ``second''
structures for the differential
operators $A$ and $B$, the structure for $B$ being taken with the opposite
sign. The Hamiltonian is the pullback of the Hamiltonian for the equation (4)
by the embedding $A,B\mapsto AB^{-1}$. Here I am going to show that their
supposition is quite right.

We have to prove that Eq.(3) can be written as $$\p_kA=(A{\d f\over \d A})_+A-
A({\d f\over \d A}A)_+,~~\p_kB=-(B{\d f\over \d B})_+B+B({\d f\over \d B}B)_+$$
where $\d f/\d A$ and $\d f/\d B$ were shown to be (see page 5):
$${\d f\over\d A}=(B^{-1}{\d f\over\d L})_-,~{\d f\over\d B}=-(B^{-1}
{\d f\over\d L}AB^{-1})_-.$$ Taking $\tf=(n/k)\int\res L^{k/n}dx$ we have
$\d f/\d L=L^{(k/n)-1}$ and $${\d f\over\d A}=(B^{-1}(AB^{-1})^{(k/n)-1})_-
,~{\d f\over\d B}=-(B^{-1}(AB^{-1})^{k/n})_-.$$ Now,
$$\p_kA=(A(B^{-1}(AB^{-1})^{(k/n)-1})_-)_+A-A(
(B^{-1}(AB^{-1})^{(k/n)-1})_-A)_+;
$$ the subscript $-$ can be omitted since if it is replaced by $+$ the outer
subscript $+$ is superfluous, and the whole expression vanishes. We have,
$$\p_kA=(AB^{-1}(AB^{-1})^{(k/n)-1})_+A-A(B^{-1}(AB^{-1})^{(k/n)-1}A)_+$$ $$=
P_kA-A((AB^{-1})^{k/n})_++A[A,B^{-1}(AB^{-1})^{( k/n)-1}]_+$$
$$=[P_k,A]+A[A,A^{-1}(AB^{-1})^{k/n}]_+=[P_k,A]+A(A^{-1}[A,(AB^{-1})^{k/n}]
)_+$$ $$=[P_k,A]+A(A^{-1}[A,(AB^{-1})_+^{k/n}])_+=[P_k,A]-A(A^{-1}[P_k,A])_+$$
which is Eq.(3a). Similarly,
$$\p_kB=(B(B^{-1}(AB^{-1})^{k/n})_-)_+B-B((B^{-1}(AB^{-1})^{k/n})_-B)_+$$
$$=(BB^{-1}(AB^{-1})^{k/n})_+B-B(B^{-1}(AB^{-1})^{k/n}B)_+=P_kB-B(B^{-1}P_kB)_+
$$ $$=[P_k,B]-B(B^{-1}[P_k,B])_+,$$ as required.

Apparently, there is the same statement about the direct sum of structures in
{\em Bonora, Liu and Xiong} [2]. It is not easy to compare since they deal not
with the equations (3) but with much more awkward equations based on the
representation (2). The problem of the first structure remains not clear.

I am very grateful to L. Feher and I. Marshall for their remark.

\end{document}